\documentclass[12pt]{article}

\title{Gravitational multi-NUT solitons, Komar masses and charges}

\global\arraycolsep=1pt
\usepackage[colorlinks=true,backref=true,linkcolor=black,anchorcolor=black,citecolor=black,filecolor=black,menucolor=black,pagecolor=black,urlcolor=black]{hyperref}
\usepackage{amsmath}
\usepackage{mathrsfs}
\usepackage{bbm}
\usepackage{dsfont}
\usepackage{amssymb}
\usepackage{textcomp}
\usepackage{caption}
\usepackage{graphicx}

\newcommand{\scal}[1]{\bigl ({#1} \bigr )}
\def\bea{\begin{eqnarray}}
\def\eea{\end{eqnarray}}
\def\be{\begin{equation}}
\def\ee{\end{equation}}
\def\ie{{\it i.e.}\ }
\newcommand{\CR}{\nonumber \\*}

\DeclareMathAlphabet{\mathpzc}{OT1}{pzc}{m}{it}

\def\r{{\tilde{r} }}

\def\L{{\cal L}}

\newcommand{\sfrac}[2]{{\scriptstyle \frac{#1}{#2}}}

\begin{document}
\allowdisplaybreaks[1]
\renewcommand{\thefootnote}{\fnsymbol{footnote}}
\def\corr{$\spadesuit $}
\def\trefle{$\clubsuit$}
\begin{titlepage}
\begin{flushright}
\
\vskip -2.5cm
{\small AEI-2008-080}\\
{\small Imperial/TP/08/KSS/01}\\
{\small CERN-PH-TH/2008-202}
\vskip 2.5cm
\end{flushright}
\begin{center}
{\Large \bf
Gravitational multi-NUT solitons,\\
\vskip 2mm
Komar masses and charges}
\\
\lineskip .75em
\vskip 3em
\normalsize
{\large  Guillaume Bossard\footnote{email address: bossard@aei.mpg.de},
Hermann Nicolai\footnote{email address: Hermann.Nicolai@aei.mpg.de} and
K.S. Stelle\footnote{email address: k.stelle@imperial.ac.uk}}\\
\vskip 1 em
$^{\ast\dagger\ddagger}${\it AEI, Max-Planck-Institut f\"{u}r Gravitationsphysik\\
Am M\"{u}hlenberg 1, D-14476 Potsdam, Germany}
\\
\vskip 1 em
$^{\ddagger}${\it Theoretical Physics Group, Imperial College London\\
Prince Consort Road, London SW7 2AZ, UK}
\\
\vskip 1 em
$^{\ddagger}${\it Theory Division, Physics Department, CERN\\
CH-1211 Geneva 23, Switzerland}

\vskip 1 em
\end{center}
\vskip 1 em
\begin{abstract}
Generalising expressions given by Komar, we give precise definitions of gravitational mass and solitonic NUT charge and we apply these to the description of a class of Minkowski-signature multi-Taub--NUT solutions without rod singularities. A Wick rotation then yields the corresponding class of Euclidean-signature gravitational multi-instantons.
\end{abstract}

\end{titlepage}
\renewcommand{\thefootnote}{\arabic{footnote}}
\setcounter{footnote}{0}


\section{Introduction}
In many respects, the Taub--NUT solution \cite{NUT} appears to be dual to the Schwarzschild solution in a fashion similar to the way a magnetic monopole is the dual of an electric charge in Maxwell theory. The Taub--NUT space-time admits closed time-like geodesics \cite{Misner} and, moreover, its analytic extension beyond the horizon turns out to be non Hausdorff \cite{Hawking}. The horizon covers an orbifold singularity which is homeomorphic to a two-sphere, although the Riemann tensor is bounded in its vicinity. These pathologies lead to the view that the Taub--NUT solution is not physical. 

Nonetheless, despite the fact that no magnetic monopole has yet been found in our universe, such magnetic dual solutions play an important r\^ole in quantum electrodynamics and especially in its non-abelian generalisation, namely Yang--Mills theory. Moreover, the stationary solutions of the Maxwell--Einstein equations admit a non-linearly realised $SU(2,1)$ symmetry group \cite{Kinnersley} which generalises the Ehlers group and which mixes together the electromagnetic and the gravity degrees of freedom. This generalises to a large class of theories, and in particular to ones that can be embedded into supergravity theories. Despite the fact that this has not been proven so far, these symmetry groups are believed to act on the non-stationary solutions as well. The major difficulty in formulating such symmetries comes from the fact that Einstein's theory is highly non-linear and consequently its dualities are poorly understood beyond the linearised level. The aim of this letter is to understand more closely the duality relations within Einstein theory by exhibiting their similarities with the example of Maxwell theory, and more specifically the similarities between NUT sources and magnetic monopoles. 

While a magnetic charge can be expressed in terms of a current associated to a vector field dual to the standard Maxwell potential, its expression as a function of the standard vector potential corresponds to a topological invariant of the associated fibre-bundle geometry. In this letter, we define the NUT charge, in a similar way, as a topological invariant associated to time-like three-cycles. We also generalise the Komar mass to the case where there is no space-like slice with compact boundary in the asymptotic region. These definitions involve a fibre-bundle construction which is very reminiscent of the one appearing in Maxwell theory. In this case the $U(1)$ fibres are orbits of the time-like isometry. 

We exhibit the similarities between the Komar NUT charge and magnetic charge through a consideration of explicit solutions involving several NUT sources. Indeed, we will give an infinite set of new regular solutions of the Einstein equations with an arbitrary odd number of NUT sources. We obtain these by acting with the $U(1)$ duality group for stationary solutions on multi-black hole solutions with both negative and positive masses. We define a coordinate patch that permits us to avoid Dirac--Misner string singularities. Then we show how one can avoid the conical singularities usually appearing in multi-black hole solutions by choosing adjacent NUT charges to be opposite in sign. Since both negative and positive NUT charge singularities are covered by horizons, these solutions define space-times which have no more (albeit also no less) pathology than the ordinary Taub--NUT space-time. 

The resolution of the Dirac--Misner string singularities requires the quantisation of NUT charge. The Chern class of an associated fibre-bundle geometry is understood to count the relative number of fundamental NUT charges of a given spacetime. The timelike three-cycles surrounding several NUT charges turn out to be diffeomorphic to Lens spaces $L(|N|,1) \cong S^3 / \mathds{Z}_{|N|}$, where $N$ is the relative number of fundamental NUT charges that lie inside the interior of the corresponding three-cycle.

We discuss in a final section the Euclidean analogues of these multi-NUT solutions which are slight generalisations of the instantons described in \cite{Gibbons}

\section{Komar NUT charge}
A.~Komar defined the mass for asymptotically Minkowski solutions of the Einstein equations through an integral over the boundary of an asymptotically space-like hypersurface $V$ in spacetime \cite{Komar}. Given an asymptotically Killing time-like vector $\kappa = \kappa^\mu \partial_\mu $, the metric permits one to define the $1$-form $g(\kappa) \equiv g_{\mu\nu} \kappa^\mu dx^\nu$, and the Komar mass is then given as a function of the $2$-form $K \equiv d g(\kappa)$ by\footnote{Komar proved in \cite{positive} that if $\kappa$  is chosen to be orthogonal to a family of minimal hypersurfaces, then the Komar mass will be positive if $V$ is chosen to be one of these hypersurfaces. However, Misner then showed in \cite{dispositive} that this prescription is either inconsistent or impossible to achieve in some relevant examples. Here we will not insist on this orthogonality prescription and the Komar mass consequently will not be necessarily positive.}
\be m \equiv \frac{1}{8 \pi }Ê\int_{\partial V} \star K \ee
where $\star$ is the Hodge star operator. Comparing then this formula to the ones defining the electric and the magnetic charges 
\be q \equiv \frac{1}{2 \pi }Ê\int_{\partial V} \star F \hspace{10mm} p \equiv \frac{1}{2 \pi }Ê\int_{\partial V}  F\ee
it seems natural to define the dual mass as the integral 
\be n \equiv \frac{1}{8 \pi }Ê\int_{\partial V}  K \ . \ee
However this integral is trivially zero because of Stokes theorem, as also would na\"{\i}vely be the one defining magnetic charge. Nevertheless, not all asymptotically flat space-times admit a well-defined asymptotically space-like hypersurface. This is the case for instance for the Taub--NUT space-time, for which the $r = {\rm const}$ slices of any space-like hypersurfaces are not closed manifolds \cite{Misner}. 

 
Let $M$ be an asymptotically flat space-time. Strictly speaking, we assume that $M$ admits a function $r$ which goes to infinity at spatial infinity and which defines a proper distance in this limit, $g^{\mu\nu}Ê\partial_\mu r \partial_\nu r \rightarrow 1$, and we assume that all the components of the Riemann tensor in any vierbein frame go to zero as $\mathcal{O}(r^{-3})$ as $r \rightarrow + \infty$. We consider stationary solutions; $\kappa$ is then a Killing vector and the second-order components of the Einstein equations can be written as
\be d \star K = 2 \star dx^\mu R_{\mu\nu} \kappa^\nu = 16 \pi G   \star dx^\mu \scal{ T_{\mu\nu} - \sfrac{1}{2} g_{\mu\nu} T } \kappa^\nu  \label{Max} 
\ee
which is very similar to the Maxwell equation. We choose the function $r$ to be invariant under the action of the time-like isometry, and choose the squared norm of the time-like Killing vector $g_{\mu\nu} \kappa^\mu \kappa^\nu \equiv - H$ to tend to $-1 + \mathcal{O}(r^{-1})$ as $r \rightarrow + \infty$. We assume that the action of the time-like isometry is free and proper on the domain of $M$ where the function $H$ is positively defined. This implies that $M$ admits an Abelian principal bundle structure over a Riemannian three-fold $V$ on this domain. 

If this principal bundle is trivial, it admits a global section $s$ which defines an embedding of $V$ into a space-like hypersurface of $M$ (outside the zeros of $H$). Otherwise it only admits a patch of local sections defined on each open set of an atlas of $V$, which we denote collectively by $s$ as well. A natural generalisation of the Komar mass formula thus consists in defining it as the integral of the pull back $s^* \star K$ of the $2$-form $\star K$  over $\partial V$. In order for this integral not to depend on the local trivialisation, $\star K$ must be horizontal and invariant in the asymptotic region. It is trivially invariant since it is built from the metric and the Killing vector, and the horizontality condition is given by asymptotic hypersurface orthogonality, \ie $i_\kappa \star K \rightarrow 0$  as $r \rightarrow + \infty$. Because of equation (\ref{Max}), $d \, i_\kappa \star K = 0$ in the vacuum and if space-time is simply-connected, there exists a function $B$ such that $ i_\kappa \star K = d B$. The horizontality condition for $\star K$ in the asymptotic region is then equivalent to the fact that $B$ tends to zero as $\mathcal{O}(r^{-1})$ as $r \rightarrow + \infty$. The $2$-form $K$ is also trivially invariant and its horizontality condition $i_\kappa K = d H \rightarrow 0 $ is satisfied because the function $H$ tends to unity as $r \rightarrow + \infty$.


We accordingly define the mass $m$ and its dual, the NUT charge $n$, by the following integrals
 \be m \equiv \frac{1}{8 \pi }Ê\int_{\partial V}  s^* \star K  \hspace{10mm}  n \equiv \frac{1}{8 \pi }Ê\int_{\partial V} s^* K \label{def}Ê\ . \ee
 
By construction, the 1-form $g(\kappa)$ is invariant under the action of the time-like isometry, and since $i_\kappa g(\kappa) = -H \rightarrow -1 $ as $r \rightarrow + \infty$, it defines a connection on the principal bundle in the asymptotic region. The NUT charge is proportional to the Chern class of the principal bundle over $\partial V$, and is thus non-zero only in the case where the latter is non-trivial. Real line bundles over a compact surface always have a vanishing Chern class, and a non-zero NUT charge implies therefore that time-like orbits are compact. 


One defines electric and magnetic charges in the same way by requiring both the Maxwell potential and its dual to be invariant under the covariant action of the time-like isometry in the asymptotic region, \ie $i_\kappa F \sim i_\kappa \star F \sim \mathcal{O}(r^{-2})$. The whole construction can be generalised to non-stationary space-times, as long as $\L_\kappa g_{\mu\nu}$ tends sufficiently fast to zero as $r \rightarrow + \infty$.


Let us now express the mass and its dual in a more explicit way.  We choose coordinates for which $\kappa = \partial_t$, in such a way that the metric is given as follows 
\be ds^2 = - H \scal{Êdt + \hat{B}_i dx^i }^2 + H^{-1} \gamma_{ij} \, dx^i dx^j \ . \ee
The vacuum Einstein equations then give $d \, i_\kappa \star K = 0$ and
\be i_\kappa \star K = - H^2 \sqrt{\gamma} {\varepsilon_i}^{jk}  \partial_j \hat{B}_k \, dx^i  =  dx^i \partial_i B  \label{Duality3} \ee
where Latin indices are raised and lowered with the three-dimensional metric $\gamma_{ij}$. The asymptotic horizontality conditions for $K$ and $\star K$ are satisfied if 
\be H = 1 - \frac{2m}{r}Ê+ \mathcal{O} (r^{-2}) \hspace{10mm} B = - \frac{2n}{r} +  \mathcal{O} (r^{-2})\ . \label{asympt}  \ee
Then $\star K$ and $K$ have the following behaviour in the asymptotic region  
\be \star K \sim \frac{\sqrt{\gamma}}{2}  {\varepsilon_{ij}}^{k} H^{-1} \partial_k H dx^i \wedge dx^j \hspace{10mm}ÊK \sim  - \partial_i \hat{B}_j \, dx^i \wedge dx^j \ . \ee
If we assume furthermore that $\gamma_{ij}$ is asymptotically Euclidean, one may verify that the parameters $m$ and $n$ appearing in (\ref{asympt}) are truly the mass and NUT charges defined by (\ref{def}). 

One obtains $s^* K =  2 n \sin \theta d\theta \wedge  d \varphi$ in polar coordinates on $\partial V \cong S^2$. $\hat{B}_i$ is thus only globally defined up to a constant, and the time coordinate also is not globally defined over the two-sphere. One defines $t_+$ and $t_-$ on the north and the south pole of the two-sphere respectively. These coordinates are related by 
\be t_+ = t_- - 4 n \varphi \ .
\ee 
Since $\varphi$ is a periodic coordinate of period $2\pi$, the time coordinates $t_\pm$ must be periodic of period $8 \pi n_0$, such that $N= \frac{n}{n_0}$ is an integer. The integer $N$ parametrizes the Chern class of the principal bundle over $S^2$, and the $r = {\rm const} $ slices for $r$ sufficiently large are diffeomorphic to the Lens space $S^3 / \mathds{Z}_{|N|}$. 

The Lens spaces are usually studied as Riemannian three-folds, but they also admit a pseudo-Riemannian metric, as does any $U(1)$ principal bundle over a Riemannian manifold. If we define a connection $\omega$ on to the principal bundle, as well as the pull-back of the metric $\gamma$ on the base by the bundle projection $\pi$, then $\omega \otimes \omega + \pi^* \gamma$ gives a natural Riemannian metric on the principal bundle, and $ -  \omega \otimes \omega + \pi^* \gamma$ a natural pseudo-Riemannian metric.

  
S.~Ramaswamy and A.~Sen obtained a similar result in \cite{Bondi}, where they defined the NUT charge as a dual of the Bondi mass instead of the Komar mass. The Bondi mass and its dual are defined using integrals involving respectively the Weyl tensor and its Hodge dual. 
  
  
The $U(1)$ principal bundles over $S^2$ are classified by their first Chern class, which is unity in the case of the Hopf fibration of $S^3$. By analogy with the case of the Maxwell theory for which the Chern class determines the relative number of fundamental Dirac monopoles, we will wish to interpret this integer as the relative number of fundamental NUT sources in General Relativity. This interpretation turns out to be right, as we shall see in the following. 
  
 \section{Multi-Taub--NUT solutions}
 We now want to consider axisymmetric stationary solutions of the Einstein equations with several NUT sources on the axial symmetry axis. We use Weyl coordinates in which
 \be  ds^2 = H^{-1} e^{2\sigma} \scal{Êdz^2 + d \rho^2} + \rho^2 H^{-1} d \varphi^2 - H ( dt + \hat{B} d \varphi ) ^2 \ . \label{Weyl}\ee  
For example, in the case of the Taub--NUT solution of mass $m$ and NUT charge $n$, the Weyl coordinates are related to the Schwarzschild ones by\footnote{Note that the radius $r$ that is commonly introduced in Weyl coordinates is not the Schwarzschild radius $\r$, but is related to it by $r = \r - m$.}
\be \rho = \sqrt{ \r^2 -2 m \r - n^2  } \sin \theta \hspace{10mm} z = (\r - m )  \cos \theta \ , \ee
in terms of which the metric is
\be ds^2 = - H \scal{ dt_\pm +  2 n ( \pm 1 -  \cos \theta ) d \varphi  }^2 + H^{-1} d\r^2 + \scal{Ê\r^2 + n^2 } \scal{Êd \theta^2 + \sin^2 \theta d \varphi^2 } \ee
with
\be H = \frac{ \r^2 - 2 m \r - n^2}{ \r^2 + n^2 } \ . \ee
In Weyl coordinates, equation (\ref{Duality3}) reduces to 
\be    \rho^{-1} H^2 \partial_\rho \hat{B} = - \partial_z B  \hspace{10mm}  \rho^{-1} H^2 \partial_z \hat{B} = \partial_\rho B \label{duality} \ee
and $B$ is the imaginary part of the so-called Ernst potential, $\mathcal{ E}Ê\equiv H + i B$. This latter satisfies the Ernst equation
\be \scal{{\cal E}  + {\cal E}^* } \left( {\partial_z}^2 + {\partial_\rho}^2 + \frac{1}{\rho} \partial_\rho \right) \mathcal{E} =  2 \partial_z {\cal E}  \partial_z {\cal E}  + 2 \partial_\rho {\cal E}  \partial_\rho {\cal E}\ . \ee
For static solutions, the Ernst potential is real and the Ernst equation reduces to the linear differential equation
\be \left(Ê {\partial_z}^2 + {\partial_\rho}^2 + \frac{1}{\rho} \partial_\rho \right) \ln \mathcal{E} = 0 \ .
\ee
The product of several real Ernst potentials thus gives a new solution. This permits one to obtain the Ernst potential of multi-black holes solutions as 
\be \mathcal{ E}Ê= \prod_{i=1}^h \frac{ r_i - c_i}{r_i + c_i} \label{multiBH} \ee 
where $2 r_i = r_{i\, +} + r_{i\, -}$ with
\be  r_{i \, \pm}  \equiv  \sqrt{ (z - z_i \pm  c_i)^2  + \rho^2} \ee
and where $z_i$ and $c_i$ define respectively the position and the (possibly negative) mass of each of the $h$ black holes. When all masses are positive, these solutions are always known to suffer from  conical singularities unless one considers an infinite chain of black holes \cite{Hermann}.


A nice way to interpret the NUT charge as a dual mass comes from the fact that the stationary solutions of Einstein's equations admit a nonlinearly realised $U(1)$ Ehlers symmetry \cite{Ehlers} which rotates the mass into the  NUT charge in the case of the Taub--NUT solutions. This $U(1)$ acts trivially on the conformal factor $\sigma$ and modifies the Ernst potential as follows
 \be \mathcal{ E}(\alpha)Ê= \frac{ \cos \alpha \, \mathcal{E} - i \sin \alpha }{\cos \alpha  - i \sin \alpha \,  \mathcal{E}} \label{so2} \ . \ee
Acting this way on the Ernst potential (\ref{multiBH}), one gets 
 \be \mathcal{E}Ê= \frac{ \cos \alpha \prod (r_i - c_i) - i \sin \alpha  \prod (r_i + c_i) }{ \cos \alpha  \prod (r_i + c_i) -i \sin \alpha   \prod (r_i - c_i) } \ee
where $ e^{2 i \alpha } c_i  = m_i + i n_i $. We then derive the potentials for the metric\footnote{\label{steps} To derive the formula for $\hat{B}$ we observe that $ H^{-2} d B = - 2 \sum  \frac{n_i dr_i}{{r_i}^2 - {c_i}^2} $, and we make use of the identities $\frac{\rho^2}{{r_i}^2 - {c_i}^2} + \frac{ ( z - z_i)^2}{{r_i}^2} = 1$ and $r_{i\, \pm}  = r_i \pm c_i \frac{z-z_i}{r_i} $ to show that 
 \be \frac{ \rho  \, \partial_z r_i}{{r_i}^2 - {c_i}^2} = - \partial_\rho \frac{z-z_i}{r_i} \hspace{10mm}  \frac{ \rho  \, \partial_\rho r_i}{{r_i}^2 - {c_i}^2} = \partial_z \frac{z-z_i}{r_i} \ .\nonumber \ee}
\be H = \frac{  \prod ( {r_i}^2 - {c_i}^2 ) }{ \cos^2 \alpha \prod (r_i + c_i)^2 + \sin^2 \alpha \prod (r_i - c_i)^2} \hspace{10mm}Ê\hat{B} = b - 2  \sum_{i=1}^h n_i \frac{z - z_i}{r_i}  \ee
 where $b$ is an undetermined integration constant coming from the duality relation (\ref{duality}). Note that the potential $\hat{B}$ is a sum of potentials for ordinary Taub--NUT solutions individually centred at $z_i$. There is one horizon on each segment $\rho = 0$,  $z_i - |c_i| \le z \le z_i + |c_i| $.  Let us consider that they are all separated, \ie that 
 \be z_{i-1} + |c_{i-1}| < z_{i} - |c_{i}| \ . \ee
 Between each adjacent pair of horizons, there is a Dirac--Misner string singularity related to the fact that the $1$-form  $d \varphi$ diverges on the symmetry axis $\rho = 0$. The Dirac--Misner string singularities are located on $h+1$ segments $D_i$, on which $\rho = 0$ and $z_{i-1} + |c_{i-1}| \le z \le z_{i} - |c_{i}|$, where we understand $-\infty  < z  \le z_1 - |c_1|$ and $z_h + |c_h|  \le z < + \infty $ for $D_1$ and $D_{h+1}$ respectively. In order to avoid such a singularity, the potential $\hat{B}$ must vanish identically on each of these segments. On the segment $D_i$, $r_j = z - z_j $ for $j < i$ and $r_j = -z + z_j $ for $j \ge i$, so one has 
 \be  \hat{B}_{|D_i} = b_i - 2 \sum_{j=1}^{i-1}Ên_j + 2 \sum_{j = i}^{h}Ên_j = 0 \ . \ee
Exactly in the same way as for the ordinary Taub--NUT solution \cite{Misner}, in order to avoid Dirac--Misner string singularities, one must define $h+1$ open sets $U_i$, such that $\cup_{i=1}^{h+1} U_i$ covers space-time outside the horizons. We define each $U_i$ as the complement of the domain $\cup_{j \ne i} D_j$ in $M$. At the intersection between $U_i$ and $U_{i+1}$, the corresponding time coordinates are related by  
\be t_{i+1} = t_i - 4 n_i \varphi  \ee 
and $\hat{B}$ is given by 
\be \hat{B}_{|U_i} = 2 \sum_{j=1}^{i-1}Ên_j - 2 \sum_{j = i}^{h}Ên_j  - 2  \sum_{j=1}^h n_j \frac{z - z_j}{r_j}  \ee
on $U_i$, in such a way that $d t_i + \hat{B}_{|U_i} d \varphi$ is globally defined on $M$.

Since $\varphi$ is a periodic coordinate, $\varphi \approx \varphi + 2\pi$, consistency requires the time coordinate also to be periodic, that is  $t_j \approx t_j + 8 \pi n_i $ for all $n_i$. In order for the manifold to be well defined, all the NUT charges $n_i$ must thus be integral multiples of a given fundamental charge $n_0$, so $t_i \approx t_i + 8 \pi n_0$. 
 
 
 We thus conclude that, even on a purely classical level, the existence of more than one NUT charge on a manifold implies the quantisation of these charges. In fact, this quantisation already occurs in Maxwell theory if one considers that its solutions are the connections of $U(1)$ principal bundles over space-time for which the curvature verifies the equation $d \star F = 0$. Indeed, the global definition of the Maxwell connection on the principal bundle similarly requires all magnetic charges to be integral multiples of a given fundamental charge.
  
 
The one-form $\omega  \equiv \frac{1}{4 n_0} ( d t_i + \hat{B}_{|U_i} d \varphi )$ defines a connection on the $U(1)$ principal bundle over $V$. For any two-cycle of $V$ surrounding a subset $I$ of the NUT charges, one computes that associated Chern class to be 
\be N_I = \frac{1}{n_0}Ê\sum_{i \in I}Ên_i \ .
\ee
The time-like three-folds that surround the NUT charges within $I$ are thus diffeomorphic to the quotient of $S^3$ by $\mathds{Z}_{|N_I|}$ acting as a discrete subgroup of $U(1)$, yielding a Lens space. We thus interpret the Chern class $N_I$ of a two-cycle as the relative number of fundamental NUT charges inside its interior.
 
 
As for multi-black hole solutions, the multi-NUT solutions generically possess conical singularities. In order to avoid such singularities, the following function must go to unity on the symmetry axis 
\be \frac{ \partial^\mu X \partial_\mu X}{4 X }  \rightarrow 1 \label{conicDef}\ee
where $X$ is the squared norm of the axisymmetric Killing vector. In Weyl coordinates this function behaves like $e^{-2\sigma}$ as $\rho \rightarrow 0$. The condition (\ref{conicDef}) is thus  equivalent to the requirement that the function $\sigma$ tend to zero in this limit. Since $\sigma$ is invariant under the duality transformation (\ref{Weyl}), one can simply compute it for the multi-black hole solutions. One gets, as a direct generalisation of the case of two positive mass black holes given in \cite{equili}, that 
\be 2 \sigma = \sum_{i=1}^h \ln \frac{ {r_i}^2 - {c_i}^2}{r_{i\, +} r_{i\, -}}  + \sum_{i < j} \ln \frac{ E_{\, i\ j}^{+-} E_{\, i\ j}^{-+} }{ E_{\, i\ j}^{--} E_{\, i\ j}^{++}}   \ee
where
\be  E_{\, i\ j}^{\pm\pm} = r_{i\, \pm} r_{j\, \pm} + ( z - z_i \pm c_i ) ( z - z_j \pm c_j ) + \rho^2 \ . \ee
On the segment $D_{k}$, the function $\sigma$ is thus constant and is equal to 
\bea \sigma_{| D_{k}} &=& \sum_{i= 1}^{k-1} \sum_{ j= k}^h \ln \frac{ (z_i - z_j)^2 - (c_i + c_j)^2 }{(z_i - z_j)^2 - (c_i - c_j)^2 } \CR
&=& \sum_{i= 1}^{k-1} \sum_{ j= k}^h {\rm sign} \, (c_i c_j) \ln \left( 1 - \frac{ 4 | c_i | [c_j| }{ \scal{ 2 |c_i | +L_{ij} }\scal{ 2 | c_j | + L_{ij} }} \right) \eea
where $L_{ij} \equiv | z_i - z_j| - |c_i| - |c_j| $ is the distance between the two horizons of the black holes  centred at $z = z_i$ and $z= z_j$ respectively. Since we require the horizons not to overlap, all the $L_{ij}$ are strictly positive and one sees that $\sigma$ can only be zero on each segment $D_k$ if some of the masses $c_i$ are negative. 

Our multi-NUT solution defines thus a perfectly smooth space-time outside the horizons if and only if
\be \prod_{i= 1}^{k-1} \prod_{j = k}^h \frac{ (z_i - z_j)^2 - (c_i + c_j)^2 }{(z_i - z_j)^2 - (c_i - c_j)^2 }  =   1\label{conic} \ee
for all $k$ between $2$ and $h$. These $h-1$ equations determine the relative positions of the NUT sources as functions of their charges.  

\section{Some examples}

Let us consider a simple class of examples with three NUT sources, with charges $n_1 = n_3 = p n_0$ and $n_2 = - q n_0$ for two integers $p$ and $q$. We also fix $z_3 = - z_2 = z_0$ and $z_1 = 0$. The absence of a conical singularity requires that
\be \frac{ \scal{ {z_0}^2 - (p- q)^2 { n_0}^2 } \scal{Ê(2 z_0 ) ^2 - ( 2 p n_0 )^2 }}{Ê \scal{ {z_0}^2 - (p + q)^2 {n_0}^2 }  ( 2 z_0 )^2 } = 1 \ . \ee
This equation can be solved for $p > 4 q$, by
\be z_0 = \frac{ p - q }{\sqrt{ 1 - \frac{4q}{p} }} n_0  \label{posi}\ee
and the horizons are disjoint for any value of $p$ and $q$.

The asymptotic $r= {\rm const }$ slices are then diffeomorphic to a Lens space $S^3 / \mathds{Z}_{N}$ with $N \equiv  2 p- q  $. The value of the Chern class in these examples is $N= 2 r + 7 q$ for strictly positive integers $r$ and $q$. 

Since $\frac{1}{4n_0} \hat{B} d \varphi $ defines the pullback of a $U(1)$ connection on each $U_i$ of $V$, we can compute the Chern class of the two-cycles in $V$ from it. We define the partitions of various two-cycles over the atlas of $V$ as depicted in the following figures\\
\vskip 1mm
\hspace{-13mm}
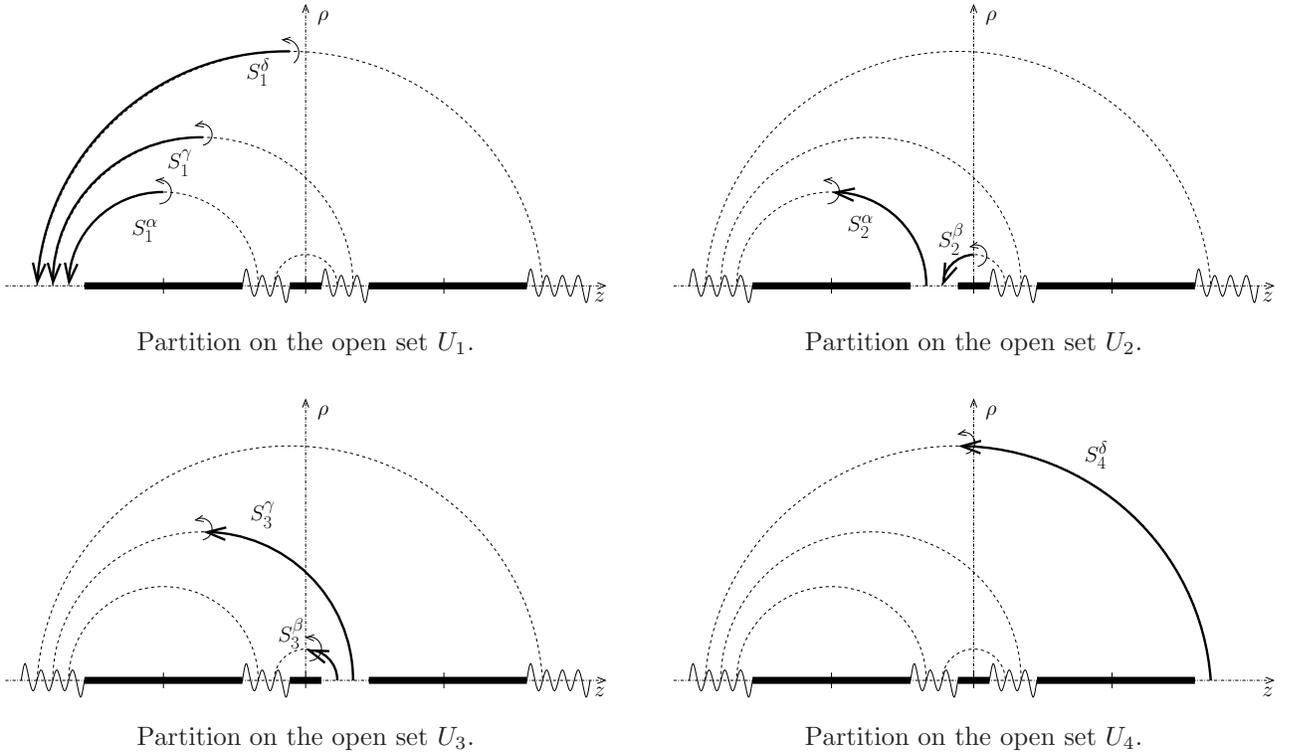
\begin{minipage}[t]{8.5cm}
\begin{center}
\resizebox{8cm}{4cm}{\begin{picture}(0,0)%
\includegraphics{fig1.pstex}%
\end{picture}%
%
%
\setlength{\unitlength}{3947sp}%
\begingroup\makeatletter\ifx\SetFigFont\undefined%
\gdef\SetFigFont#1#2#3#4#5{%
  \reset@font\fontsize{#1}{#2pt}%
  \fontfamily{#3}\fontseries{#4}\fontshape{#5}%
  \selectfont}%
\fi\endgroup%
\begin{picture}(11424,5770)(289,481)
\put(6226,5939){\makebox(0,0)[lb]{\smash{{\SetFigFont{12}{14.4}{\familydefault}{\mddefault}{\updefault}{\color[rgb]{0,0,0}\Huge  $\rho$}%
}}}}
\put(4876,4814){\makebox(0,0)[lb]{\smash{{\SetFigFont{12}{14.4}{\familydefault}{\mddefault}{\updefault}{\color[rgb]{0,0,0}\Huge  $S^ \delta_1$}%
}}}}
\put(11476,539){\makebox(0,0)[lb]{\smash{{\SetFigFont{12}{14.4}{\familydefault}{\mddefault}{\updefault}{\color[rgb]{0,0,0}\Huge $z$}%
}}}}
\put(3376,3089){\makebox(0,0)[lb]{\smash{{\SetFigFont{12}{14.4}{\familydefault}{\mddefault}{\updefault}{\color[rgb]{0,0,0}\Huge $S^ \gamma_1$}%
}}}}
\put(2701,1814){\makebox(0,0)[lb]{\smash{{\SetFigFont{12}{14.4}{\familydefault}{\mddefault}{\updefault}{\color[rgb]{0,0,0}\Huge $S^ \alpha_1$}%
}}}}
\end{picture}%
}
 {\footnotesize Partition on the open set $U_1$.}
\end{center}
\end{minipage}
\begin{minipage}[t]{0.2cm}
\hspace{5mm}
\end{minipage}
\begin{minipage}[t]{8.5cm}
\begin{center}
\resizebox{8cm}{4cm}{\begin{picture}(0,0)%
\includegraphics{fig2.pstex}%
\end{picture}%
%
%
\setlength{\unitlength}{3947sp}%
\begingroup\makeatletter\ifx\SetFigFont\undefined%
\gdef\SetFigFont#1#2#3#4#5{%
  \reset@font\fontsize{#1}{#2pt}%
  \fontfamily{#3}\fontseries{#4}\fontshape{#5}%
  \selectfont}%
\fi\endgroup%
\begin{picture}(11424,5770)(289,481)
\put(6226,5939){\makebox(0,0)[lb]{\smash{{\SetFigFont{12}{14.4}{\familydefault}{\mddefault}{\updefault}{\color[rgb]{0,0,0}\Huge  $\rho$}%
}}}}
\put(11476,539){\makebox(0,0)[lb]{\smash{{\SetFigFont{12}{14.4}{\familydefault}{\mddefault}{\updefault}{\color[rgb]{0,0,0}\Huge $z$}%
}}}}
\put(3601,1889){\makebox(0,0)[lb]{\smash{{\SetFigFont{12}{14.4}{\familydefault}{\mddefault}{\updefault}{\color[rgb]{0,0,0}\Huge $S^\alpha_2$}%
}}}}
\put(5326,1589){\makebox(0,0)[lb]{\smash{{\SetFigFont{12}{14.4}{\familydefault}{\mddefault}{\updefault}{\color[rgb]{0,0,0}\Huge $S^ \beta_2$}%
}}}}
\end{picture}%
}
 {\footnotesize Partition on the open set $U_2$.}
\end{center}
\end{minipage} \\
\vskip 5mm
\hspace{-13mm}
\begin{minipage}[t]{8.5cm}
\begin{center}
\resizebox{8cm}{4cm}{\begin{picture}(0,0)%
\includegraphics{fig3.pstex}%
\end{picture}%
%
%
\setlength{\unitlength}{3947sp}%
\begingroup\makeatletter\ifx\SetFigFont\undefined%
\gdef\SetFigFont#1#2#3#4#5{%
  \reset@font\fontsize{#1}{#2pt}%
  \fontfamily{#3}\fontseries{#4}\fontshape{#5}%
  \selectfont}%
\fi\endgroup%
\begin{picture}(11424,5770)(289,481)
\put(6226,5939){\makebox(0,0)[lb]{\smash{{\SetFigFont{12}{14.4}{\familydefault}{\mddefault}{\updefault}{\color[rgb]{0,0,0}\Huge  $\rho$}%
}}}}
\put(11476,539){\makebox(0,0)[lb]{\smash{{\SetFigFont{12}{14.4}{\familydefault}{\mddefault}{\updefault}{\color[rgb]{0,0,0}\Huge $z$}%
}}}}
\put(5476,1589){\makebox(0,0)[lb]{\smash{{\SetFigFont{12}{14.4}{\familydefault}{\mddefault}{\updefault}{\color[rgb]{0,0,0}\Huge $S^\beta_3$}%
}}}}
\put(4951,3914){\makebox(0,0)[lb]{\smash{{\SetFigFont{12}{14.4}{\familydefault}{\mddefault}{\updefault}{\color[rgb]{0,0,0}\Huge $S^\gamma_3$}%
}}}}
\end{picture}%
}
 {\footnotesize Partition on the open set $U_3$.}
\end{center}
\end{minipage}
\begin{minipage}[t]{0.2cm}
\hspace{5mm}
\end{minipage}
\begin{minipage}[t]{8.5cm}
\begin{center}
\resizebox{8cm}{4cm}{\begin{picture}(0,0)%
\includegraphics{fig4.pstex}%
\end{picture}%
%
%
\setlength{\unitlength}{3947sp}%
\begingroup\makeatletter\ifx\SetFigFont\undefined%
\gdef\SetFigFont#1#2#3#4#5{%
  \reset@font\fontsize{#1}{#2pt}%
  \fontfamily{#3}\fontseries{#4}\fontshape{#5}%
  \selectfont}%
\fi\endgroup%
\begin{picture}(11424,5770)(289,481)
\put(6226,5939){\makebox(0,0)[lb]{\smash{{\SetFigFont{12}{14.4}{\familydefault}{\mddefault}{\updefault}{\color[rgb]{0,0,0}\Huge  $\rho$}%
}}}}
\put(11476,539){\makebox(0,0)[lb]{\smash{{\SetFigFont{12}{14.4}{\familydefault}{\mddefault}{\updefault}{\color[rgb]{0,0,0}\Huge $z$}%
}}}}
\put(8101,5039){\makebox(0,0)[lb]{\smash{{\SetFigFont{12}{14.4}{\familydefault}{\mddefault}{\updefault}{\color[rgb]{0,0,0}\Huge $S^\delta_4$}%
}}}}
\end{picture}%
}
 {\footnotesize Partition on the open set $U_4$.}
\end{center}
\end{minipage} \\
\vskip 5mm
The Chern class of the cycle $S^\alpha$ is given by
\be N_\alpha = \frac{1}{8 \pi n_0} \int_{S^\alpha_2} d \hat{B}_{|U_2} \wedge d \varphi  +   \frac{1}{8 \pi n_0} \int_{S^\alpha_1} d \hat{B}_{|U_1 } \wedge d \varphi  = \frac{1}{8 \pi n_0} \int_{\partial S^\alpha_2} \scal{Ê \hat{B}_{|U_2}  -  \hat{B}_{|U_1} } d \varphi = p \ee 
and one computes in the same way that $N_\beta = - q,\, N_\gamma = p - q$ and $N_\delta = 2p - q$. 

\section{Gravitational instantons}
We recall that the quantum mechanics of a particle in a Taub--NUT space-time requires the quantisation of the product of its mass with the NUT charge of space-time \cite{Henneaux}, exactly as in the case of a magnetic monopole.  Moreover, the NUT charge as defined in \cite{forever} for not necessarily stationary space-times is shown to be preserved by small deformations of the solutions through the introduction of gravitational waves. 

The Euclidean self-dual Taub--NUT solutions might play a r\^ole in quantum gravity very similar to the one played by instantons in gauge theories \cite{Eguchi}. The analogue of the instanton number would then be given by the Chern class of the asymptotic Lens space, in the sense that the action evaluated for such a solution is proportional to $|N|$. The index of the Dirac operator is however given by the Pontryagin number. 

The solitons we have described in this letter are the Minkowski analogues of the instantons described in \cite{Gibbons} with the slight generalisation of considering both positive and negative mass. However, the singularities associated with negative masses are not removed by the effects of the NUT charges in the Euclidean case. 

One can Wick rotate the Minkowskian solitons to Euclidean-signature solutions by choosing a complex pure imaginary parameter for the duality transformation (\ref{so2}). Indeed, for the Riemannian metric in Weyl coordinates,
\be  ds^2 = H^{-1} e^{2\sigma} \scal{Êdz^2 + d \rho^2} + \rho^2 H^{-1} d \varphi^2 + H ( d \psi + \hat{B} d \varphi ) ^2 \label{WeylE}\ee
the Euclidean Ernst equation is 
\be \scal{{\cal E}_+  + {\cal E}_- } \left( {\partial_z}^2 + {\partial_\rho}^2 + \frac{1}{\rho} \partial_\rho \right)\mathcal{E}_\pm =  2 \partial_z {\cal E}_\pm  \partial_z {\cal E}_\pm  + 2 \partial_\rho {\cal E}_\pm  \partial_\rho {\cal E}_\pm \ee 
where the real Ernst potentials are $\mathcal{E}_\pm \equiv H \pm B$, with $B$ derived from $\hat{B}$ using equation (\ref{duality}). For a static Ernst potential, \ie one satisfying $\mathcal{E}_+ = \mathcal{E}_-$, the Euclidean Ernst equation is identical to the Minkowski one, and the multi-black hole solutions are thus  solutions of the Euclidean theory as well. The Euclidean Ernst equation is the equation of motion of an $SL(2,\mathds{R})/ SO(1,1)$ non-linear sigma model, and it is left invariant by the $SO(1,1)$ Ehlers transformation
\be \mathcal{E}_\pm (\alpha) = \frac{Ê\cosh \alpha \, \mathcal{E}_\pm \mp \sinh \alpha }{\cosh \alpha \mp \sinh \alpha \, \mathcal{E}_\pm} \ .
\ee
Applying this transformation, we obtain the following potentials for the Riemannian metric (\ref{WeylE})\footnote{See footnote \ref{steps}.}
 \be H = \frac{  \prod ( {r_i}^2 - {c_i}^2 ) }{ \cosh^2 \alpha \prod (r_i + c_i)^2 -\sinh^2 \alpha \prod (r_i - c_i)^2} \hspace{10mm}Ê\hat{B} = b_i - 2  \sum_{i=1}^h n_i \frac{z - z_i}{r_i}  \ee
where $r_i$ is defined as in the Minkowski case and the mass and the NUT charges are given by $m_i \equiv \cosh 2 \alpha \, c_i$ and $n_i \equiv \sinh 2 \alpha \, c_i$. The resolution of the Dirac--Misner string singularities goes the same way. All the NUT charges are thus required to be integral multiples of a fundamental NUT charge $n_0$, and the imaginary time coordinate $\psi$ is again periodic, with period $8\pi n_0$.

However one can not get rid of the conical singularities in the Euclidean case without introducing singularities associated with negative masses. The only regular instantons left over are thus the single instanton with $m = \frac{5}{4} |n|$ and the self-dual instantons for which $c_i = 0$ \cite{Gibbons}. 

The (anti)self-dual gravitational instantons with $m_i = \pm n_i$ \cite{instantons} can be obtained by taking the limit $c_i \rightarrow 0$, $\alpha \rightarrow \pm \infty$ while holding $\cosh 2 \alpha \, c_i$ fixed and equal to $m_i$.  In this limit, the Ernst potentials behave as
\be \mathcal{E}_\pm = \frac{ 1 - e^{\pm 2 \alpha} \sum \frac{c_i}{r_i} }{1 + e^{\pm 2 \alpha} \sum \frac{c_i}{r_i} } + \mathcal{O}({c_i}^2) \ee
and one computes that the function $r_i$ becomes $\sqrt{ \rho^2 + (z - z_i)^2}$ and   
\be H^{-1} = 1 + 2 \sum_{i=1}^h \frac{m_i}{r_i} \ . \ee
The Ernst potentials then verify $\mathcal{E}_\mp = 1$ for $n_i = \pm m_i$ respectively, and the Ernst equation reduces to the linear differential equation
\be \left( {\partial_z}^2 + {\partial_\rho}^2 + \frac{1}{\rho} \partial_\rho \right) \scal{Ê\mathcal{E}_+ + \mathcal{E}_- }^{-1} = 0 \ .\ee
For (anti)self-dual instantons $\mathcal{E}_{\mp} = {\rm const} $, $\sigma = 0$ and equation (\ref{conic}) turns out to be satisfied independently of the position of the sources on the axis. However, the absence of Euclidean NUT singularities nevertheless requires all masses to be equal to $n_0$. 

\subsection*{Acknowledgements}
We thank J.~Bi\v{c}\'{a}k and A.~Kleinschmidt for discussions. The research of K.S.S.\ was supported in part by the EU under contract MRTN-CT-2004-005104, by the STFC under rolling grant PP/D0744X/1 and by the Alexander von Humboldt Foundation through the award of a Research Prize. K.S.S. would like to thank the Albert Einstein Institute and CERN for hospitality during the course of the work.

\end{document}